\documentclass[prl,twocolumn,showpacs]{revtex4}

\usepackage{graphicx}
\usepackage{amsmath}
\usepackage{amssymb}
\usepackage{dcolumn}
\usepackage{bm}
\usepackage{color}

\begin{document}

\title{Next-to-leading order QCD corrections to the top quark decay \\
via model-independent FCNC couplings}

\author{Jia Jun Zhang}
\author{Chong Sheng Li}
 \email{csli@pku.edu.cn}
 \author{Jun Gao}
 \author{Hao Zhang}
\author{Zhao Li}
\affiliation{Department of Physics and State Key Laboratory of
Nuclear Physics and Technology, Peking University, Beijing 100871,
China}
\author{C.-P. Yuan$^1$}
\author{Tzu-Chiang Yuan$^2$}
\affiliation{${}^1$Department of Physics $\&$ Astronomy, Michigan
State
University, East Lansing, Michigan 48824-1116, USA\\
${}^2$Institute of Physics, Academia Sinica, Nankang, Taipei 11529,
Taiwan}

\date{\today}

\begin{abstract}
D0 and CDF collaborations at the Fermilab Tevatron have searched for
non-standard-model single top-quark production and set upper limits
on the anomalous top quark flavor-changing neutral current (FCNC)
couplings $\kappa^g_{tc}/\Lambda$ and $\kappa^g_{tu}/\Lambda$ using
the measurement of total cross section calculated at the
next-to-leading order (NLO) in QCD. In this Letter, we report on the
effect of anomalous FCNC couplings to various decay branching ratios
of the top quark, calculated at the NLO. This result is not only
mandatory for a consistent treatment of both the top quark
production and decay via FCNC couplings by D0 and CDF at the
Tevatron but is also important for the study of ATLAS and CMS
sensitivity to these anomalous couplings at the CERN LHC. We find
that the NLO corrections to the partial decay widths of the three
decay channels $t\rightarrow q+g$, $t\rightarrow q+\gamma$ and
$t\rightarrow q+Z$ are at the order of 10\% in magnitude and modify
their branching ratios by about 20\%, 0.4\% and 2\%, respectively,
as compared to their leading order predictions.
\end{abstract}

\maketitle


The top quark FCNC processes $t\rightarrow q+V\quad(V=g,\gamma,Z)$
have tiny branching ratios in the standard model (SM) and are
probably unmeasurable at the CERN Large Hadron Collider (LHC) and
future colliders. On the other hand, any positive signal of these
rare decay events would definitely imply some new physics beyond the
SM. As the LHC will produce abundant top quark events (about $10^8$
per year), even in the initial low luminosity run
($\sim10\,\mathrm{fb}^{-1}/\mathrm{year}$) $8\times10^6$ top quark
pairs and $3\times 10^6$ single top quark will be produced yearly,
one may anticipate to discover first hint of new physics by
observing anomalous couplings in the top quark sector.

Recently, from their measurements of the total cross sections both
D0 \cite{Abazov:2007ev} and CDF \cite{Aaltonen:2008qr}
collaborations at the Fermilab Tevatron have searched for
non-standard-model single top-quark production and set upper limits
on the anomalous FCNC couplings $\kappa^g_{tc}/\Lambda$ and
$\kappa^g_{tu}/\Lambda$, where the leading order (LO) cross sections
have been scaled to NLO \cite{Liu:2005dp} (or resummation
\cite{Yang:2006gs}) predictions. At the LHC, ATLAS collaboration has
presented its sensitivity to studying FCNC top decays
\cite{Carvalho:2007yi}. These results show that studying top quark
FCNC couplings will provide a good probe to new physics beyond the
SM. Although there are many discussions in the literature on rare
decay processes involving model-independent top quark FCNC couplings
\cite{Han:1996ep, Han:1996ce, Obraztsov:1997if, Abe:1997fz,
Beneke:2000hk, Chikovani:2000wi, AguilarSaavedra:2004wm}, most of
them were based on LO calculations. However, especially for
$t\rightarrow q+g$, due to the large uncertainties from the
renormalization scale dependence in its LO prediction through the
strong coupling constant $\alpha_s$, it is necessary to improve the
theoretical prediction to NLO in order to match the expected
experimental accuracy at the LHC.

In this Letter, we present the analytic results of the NLO QCD
corrections to the partial decay widths and decay branching ratios
of top quark via anomalous FCNC couplings for the three processes
$t\rightarrow q+g$, $t\rightarrow q+\gamma$ and $t\rightarrow q+Z$,
which are the most commonly studied decay channels by the
experimentalists at the Tevatron and the LHC.


New physics effects involved in top quark FCNC processes can be
incorporated in a model-independent way into an effective Lagrangian
which includes the dimension-5 operators as listed below
\cite{Beneke:2000hk}
\begin{eqnarray}\label{eq1}
\mathcal{L}^{\mathrm{eff}}&=&-\frac{e}{\sin2\theta_W}\sum_{q=u,c}
\frac{\kappa^Z_{tq}}{\Lambda}\bar{q}\sigma^{\mu\nu}(f^Z_{tq}+\mathrm{i}h^Z_{tq}
\gamma_5)tZ_{\mu\nu}\notag\\
&&-e\sum_{q=u,c}\frac{\kappa^{\gamma}_{tq}}{\Lambda}\bar{q}\sigma^{\mu\nu}
(f^{\gamma}_{tq}+\mathrm{i}h^{\gamma}_{tq}\gamma_5)tA_{\mu\nu}\\
&&-g_s\sum_{q=u,c}\frac{\kappa^g_{tq}}{\Lambda}\bar{q}\sigma^{\mu\nu}T^a
(f^g_{tq}+\mathrm{i}h^g_{tq}\gamma_5)tG_{\mu\nu}^a+\mathrm{H.c}.
,\notag
\end{eqnarray}
where $\Lambda$ is the new physics scale, $\theta_W$ is the Weinberg
angle, and $T^a$ are the Gell-Mann matrices. $\kappa^V_{tq}$ with
$V=g,\gamma$ and $Z$ are normalized to be real and positive, while
$f^V_{tq}$ and $h^V_{tq}$ are complex numbers in general. Since the
partial widths in our calculation depend only on the product of
$\kappa$ and $|f|^2+|h|^2$, we could set  $|f|^2+|h|^2=1$ by
redefining the parameter $\kappa$.

From the effective Lagrangian given by Eq.~\eqref{eq1}, we obtain
the following LO partial decay width of the FCNC top decays in
$D=4-2\epsilon$ dimension,
\begin{eqnarray}
\Gamma_0^\epsilon(t\rightarrow q+g)
&=&\frac{8\alpha_sm_t^3}{3}\left(\frac{\kappa^g_{tq}}{\Lambda}\right)^2C_{\epsilon},\notag
\end{eqnarray}
\begin{eqnarray}
\Gamma_0^\epsilon(t\rightarrow q+\gamma)&=&2\alpha
m_t^3\left(\frac{\kappa^{\gamma}_{tq}}{\Lambda}\right)^2C_{\epsilon},\\
\Gamma_0^\epsilon(t\rightarrow q+Z)&=&\frac{\alpha
m_t^3\beta_Z^{4-4\epsilon}}{\sin^22\theta_W}\left(\frac{\kappa^Z_{tq}}{\Lambda}\right)^2
(3-\beta_Z^2-2\epsilon)\frac{C_{\epsilon}}{1-\epsilon},\notag
\end{eqnarray}
where the masses of light quarks $q$ ($q=u,c$) have been neglected,
$\beta_Z\equiv(1-M_Z^2/m_t^2)^{1/2}$ and
$C_{\epsilon}=\frac{\Gamma(2-\epsilon)}{\Gamma(2-2\epsilon)}\left(\frac{4\pi\mu^2}{m_t^2}\right)^\epsilon$.
Here, we also define $\Gamma_0(t\rightarrow q+V) =
\Gamma_0^\epsilon(t\rightarrow q+V)|_{\epsilon\rightarrow0}$, which
are consistent with the expressions in Ref.~\cite{Beneke:2000hk}.


Below, we present in details our calculation for the inclusive decay
width of the top quark, decaying into hadrons, up to the NLO with
the LO partonic process denoted as $t \to q + g$. The final results
of $t\rightarrow q+\gamma$ and $t\rightarrow q+Z$ are also given for
completeness.

At the NLO, we need to include both one-loop virtual gluon
corrections and real gluon emission contribution. We use
$D=4-2\epsilon$ dimensional regularization to regulate both
ultraviolet (UV) and infrared (IR) (soft and collinear) divergences
so that all divergences appear as $1/\epsilon^{\alpha}$ with
$\alpha=1$ and $2$. The UV singularities cancel after summing up the
contributions from the one-loop virtual diagrams and counterterms
according to the same convention used in Ref.~\cite{Liu:2005dp}. The
soft-singularities cancel after adding up the virtual and real
radiative corrections. To cancel collinear singularities, we need to
also include contribution induced from gluon splitting to light
quark pairs at the same order in QCD coupling.

The renormalized virtual corrections to the partial decay width of
$t\rightarrow q+g$ is
\begin{eqnarray}
\lefteqn{\Gamma_{\mathrm{virt}}^g=\frac{\alpha_s}{6\pi}\Gamma_0^\epsilon(t\rightarrow
q+g)\left\{-\frac{13}{\epsilon_{IR}^2}\right.}\notag\\
&&+\frac{1}{\epsilon_{IR}}\left[
-13\ln\frac{4\pi\mu^2}{m_t^2}+13\gamma_E+N_f-\frac{53}{2}\right]\notag\\
&&+\left[-\frac{13}{2}\left(\ln\frac{4\pi\mu^2}{m_t^2}
-\gamma_E\right)^2-12\ln\frac{\mu^2}{m_t^2}\right.\notag\\
&&\left.\left.+\left(N_f-\frac{53}{2}\right)(\ln4\pi-\gamma_E)+\frac{55\pi^2}{12}-23\right]\right\}.
\end{eqnarray}
Here $\mu$ is a renormalization scale and $N_f$ denotes the number
of active flavors at the scale $\mu$. All the ultraviolet
divergences have canceled in $\Gamma^g_{\mathrm{virt}}$, as they
must, but the infrared divergent piece is still present.

The contribution from real gluon emission ($t\rightarrow q+g+g$) is
denoted as $\Gamma_{\mathrm{real}}(t\rightarrow q+g+g)$. In order to
cancel all the collinear singularities in the sum of virtual and
real radiative corrections, we also need to include the
contributions from gluon splitting into a pair of quark and
anti-quark in the collinear region, which is denoted as
$\Gamma_{\mathrm{real}}(t\rightarrow q+q'+\bar{q}')$. Note that
there are two configurations of final state when the flavor of the
light quark coming from gluon splitting is the same as the light
quark directly from the FCNC coupling, and only one configuration
when they are different. We find that at the NLO
\begin{eqnarray}
\lefteqn{\Gamma_{\mathrm{real}}(t\rightarrow
q+g+g)=\frac{\alpha_s}{6\pi}\Gamma_0^\epsilon(t\rightarrow
q+g)\left[
\frac{13}{\epsilon_{IR}^2}\right.}\notag\\
&& +\frac{1}{\epsilon_{IR}}\left(13\ln\frac{4\pi\mu^2}{m_t^2}
-13\gamma_E+\frac{53}{2}\right)\notag\\
&&+\frac{13}{2}\left(\ln\frac{4\pi\mu^2}{m_t^2}-\gamma_E\right)^2\notag+
\frac{53}{2}\left(\ln\frac{4\pi\mu^2}{m_t^2}-\gamma_E\right)\notag\\
&&-\frac{31\pi^2}{4}+\frac{171}{2}\biggr],
\end{eqnarray}
and
\begin{eqnarray}
\lefteqn{\Gamma_{\mathrm{real}}(t\rightarrow
q+q'+\bar{q}')=\frac{\alpha_s}{6\pi}\Gamma_0^\epsilon(t\rightarrow
q+g)\left[
-\frac{1}{\epsilon_{IR}}N_f \right.}\notag\\
&&\left.+N_f\left(\gamma_E
-\ln\frac{4\pi\mu^2}{m_t^2}-3\right)-\frac{1}{12}\right].\hspace{2cm}
\end{eqnarray}
After adding them together, the total real contributions can be
written as
\begin{eqnarray}
\Gamma_{\mathrm{real}}^g&=&\Gamma_{\mathrm{real}}(t\rightarrow
q+g+g)+\Gamma_{\mathrm{real}}(t\rightarrow q+q'+\bar{q}')\notag\\
&=&\frac{\alpha_s}{6\pi}\Gamma_0^\epsilon(t\rightarrow
q+g)\left\{\frac{13}{\epsilon_{IR}^2}\right.\notag\\
&&-
\frac{1}{\epsilon_{IR}}\left[-13\ln\frac{4\pi\mu^2}{m_t^2}+13\gamma_E
+N_f-\frac{53}{2}\right]\notag\\
&&+\left[\frac{13}{2}\left(\ln\frac{4\pi\mu^2}{m_t^2}-\gamma_E\right)^2+\frac{53}{2}
\left(\ln\frac{4\pi\mu^2}{m_t^2}-\gamma_E\right)\right.\notag\\
&&\left.\left.-N_f\left(\ln\frac{4\pi\mu^2}{m_t^2}+3-\gamma_E\right)
-\frac{31\pi^2}{4}+\frac{1025}{12}\right]\right\}.\notag\\
\end{eqnarray}
Combining the real and virtual contributions, we obtain the full NLO
corrections for $t\rightarrow q+g$ as
\begin{eqnarray}\label{t2qgfinal}
\lefteqn{\Gamma_{\mathrm{NLO}}(t\rightarrow
q+g)=\Gamma_{\mathrm{virt}}^g+\Gamma_{\mathrm{real}}^g}\notag\\
&=&\frac{\alpha_s}{72\pi}\Gamma_0(t\rightarrow
q+g)\left[174\ln\left(\frac{\mu^2}{m_t^2}\right)\right.\notag\\
&&-12N_f\ln\left(\frac{\mu^2}{m_t^2}\right)-36N_f
-38\pi^2+749\biggr].
\end{eqnarray}
Note that all the infrared divergences have been canceled in
Eq.~\eqref{t2qgfinal} as they should.

As for $t\rightarrow q+\gamma$ and $t\rightarrow q+Z$, the NLO QCD
corrections to the partial decay widths are given by
\begin{eqnarray}\label{t2qgammafinal}
\lefteqn{\Gamma_{\mathrm{NLO}}(t\rightarrow
q+\gamma)=\Gamma_{\mathrm{virt}}^{\gamma}+\Gamma_{\mathrm{real}}^{\gamma}}\notag\\
&=&\frac{2\alpha_s}{9\pi}\Gamma_0(t\rightarrow
q+\gamma)\left[-3\ln\left(\frac{\mu^2}{m_t^2}\right)-2\pi^2+8\right],
\end{eqnarray}
and
\begin{eqnarray}\label{t2qZfinal}
\lefteqn{\Gamma_{\mathrm{NLO}}(t\rightarrow
q+Z)=\Gamma_{\mathrm{virt}}^Z+\Gamma_{\mathrm{real}}^Z}\notag\\
&=&\frac{\alpha_s}{3\pi}\Gamma_0(t\rightarrow
q+Z)\left[-2\ln\left(\frac{\mu^2}{m_t^2}\right)
-\frac{4(9-\beta_Z^2)}{3-\beta_Z^2}\ln\beta_Z\right.\notag\\
&&-\frac{(1-\beta_Z^2)(1+6\beta_Z^2-3\beta_Z^4)}
{\beta_Z^4(3-\beta_Z^2)}\ln(1-\beta_Z^2)\\
&&\left.+4\mathrm{Li}_2\left(-\frac{1-\beta_Z^2}{\beta_Z^2}\right)
+\frac{1+3\beta_Z^2}{\beta_Z^2(3-\beta_Z^2)}-\frac{4\pi^2}{3}
+\frac{10}{3}\right],\notag
\end{eqnarray}
respectively. Hence, up to the NLO, the partial decay widths of the
three FCNC decays can be obtained by $\Gamma(t\rightarrow
q+V)=\Gamma_0(t\rightarrow q+V)+\Gamma_{\mathrm{NLO}}(t\rightarrow
q+V)$.

In order to study the effect of NLO corrections to decay branching
ratios, we define the following branching ratio for latter numerical
analysis:
\begin{eqnarray}
BR_{\mathrm{LO}}(t\rightarrow q+V)&=&\frac{\Gamma_0(t\rightarrow
q+V)}{\Gamma_0(t\rightarrow W+b)},\\
BR_{\mathrm{NLO}}(t\rightarrow q+V)&=&\frac{\Gamma(t\rightarrow
q+V)}{\Gamma(t\rightarrow W+b)}.
\end{eqnarray}
The partial decay width for the dominant top quark decay mode of
$t\rightarrow W+b$ at the NLO can be found in Ref.~\cite{Li:1990qf},
namely,
\begin{eqnarray}
\lefteqn{\Gamma(t\rightarrow W+b)=\Gamma_0(t\rightarrow
W+b)\biggr\{1}\notag\\
&&+\frac{2\alpha_s}{3\pi}\left[2\left(\frac{(1-\beta_W^2)(2\beta_W^2-1)(\beta_W^2-2)}
{\beta_W^4(3-2\beta_W^2)}\right)\ln(1-\beta_W^2)\right.\notag\\
&&-\frac{9-4\beta_W^2}{3-2\beta_W^2}\ln\beta_W^2
+2\mathrm{Li}_2(\beta_W^2)\notag\\
&&\left.\left.-2\mathrm{Li}_2(1-\beta_W^2)-\frac{6\beta_W^4-3\beta_W^2-8}
{2\beta_W^2(3-2\beta_W^2)}-\pi^2\right]\right\},
\end{eqnarray}
with
\begin{eqnarray}
\Gamma_0(t\rightarrow
W+b)&=&\frac{G_Fm_t^3}{8\sqrt{2}\pi}|V_{tb}|^2\beta_W^4(3-2\beta_W^2),\hspace{1.5cm}
\end{eqnarray}
where $\beta_W\equiv(1-m_W^2/m_t^2)^{1/2}$.


\begin{table}
\caption{\label{table1}Branching ratios as functions of
$\kappa^V_{tq}/\Lambda$. Here $\mu=m_t$.}
\begin{ruledtabular}
\begin{tabular}{cccc}
BR [in unit of $(\frac{\kappa^V_{tq}}{\Lambda}\mathrm{TeV})^2$]
&LO&NLO&NLO/LO\\[0.5em]\hline\\
$t\rightarrow
q+g$&1.0010&1.1964&1.195\\[0.25em]
$t\rightarrow q+\gamma$&0.0544&0.0542&0.996\\[0.25em]
$t\rightarrow q+Z$&0.0448&0.0458&1.022\\[0.25em]
\end{tabular}
\end{ruledtabular}
\end{table}
\begin{table}
\caption{\label{table2}Decay widths as functions of
$\kappa^V_{tq}/\Lambda$. Here $\mu=m_t$.}
\begin{ruledtabular}
\begin{tabular}{cccc}
Width [in unit of
$(\frac{\kappa^V_{tq}}{\Lambda}\mathrm{TeV})^2\mathrm{GeV}$]
&LO&NLO&NLO/LO\\[0.5em]\hline\\ $t\rightarrow
q+g$&1.443&1.577&1.09\\[0.25em]
$t\rightarrow q+\gamma$&0.078&0.071&0.91 \\[0.25em]
$t\rightarrow q+Z$&0.065&0.060&0.93\\[0.25em]
\end{tabular}
\end{ruledtabular}
\end{table}
For the numerical calculation of the branching ratios, we take the
top quark width given in Ref.~\cite{Li:1990qf} with the following
parameters \cite{Amsler:2008zz},
\begin{eqnarray*}
&&(m_W, \, m_Z, \, m_t)=(80.398,\, 91.1876,\, 171.2)\,\mathrm{GeV},\\
&&N_f=5,\quad\alpha=1/128,\quad\sin^2\theta_W=0.231,\\
&&\qquad V_{tb}=1, \qquad
G_F=1.16637\times10^{-5}\,\mathrm{GeV^{-2}}.
\end{eqnarray*}
We analyze our results as functions of FCNC couplings
$\kappa^V_{tq}/\Lambda$ and renormalization scale $\mu$. For
$\mu=m_t$, our numerical results are listed in Table~\ref{table1}
and Table~\ref{table2} showing the NLO effect to various decay
branching ratios and partial decay widths, respectively. From
Table~\ref{table1}, we see that the NLO correction increases the LO
branching ratio by about $20\%$ for $t\rightarrow q+g$, while the
NLO corrections are much smaller for the other two decay modes. But
we note that the NLO results modify the LO widths by about $10\%$ in
magnitude for all the three modes.

\begin{figure}
\scalebox{0.63}{\includegraphics*[120,420][490,655]{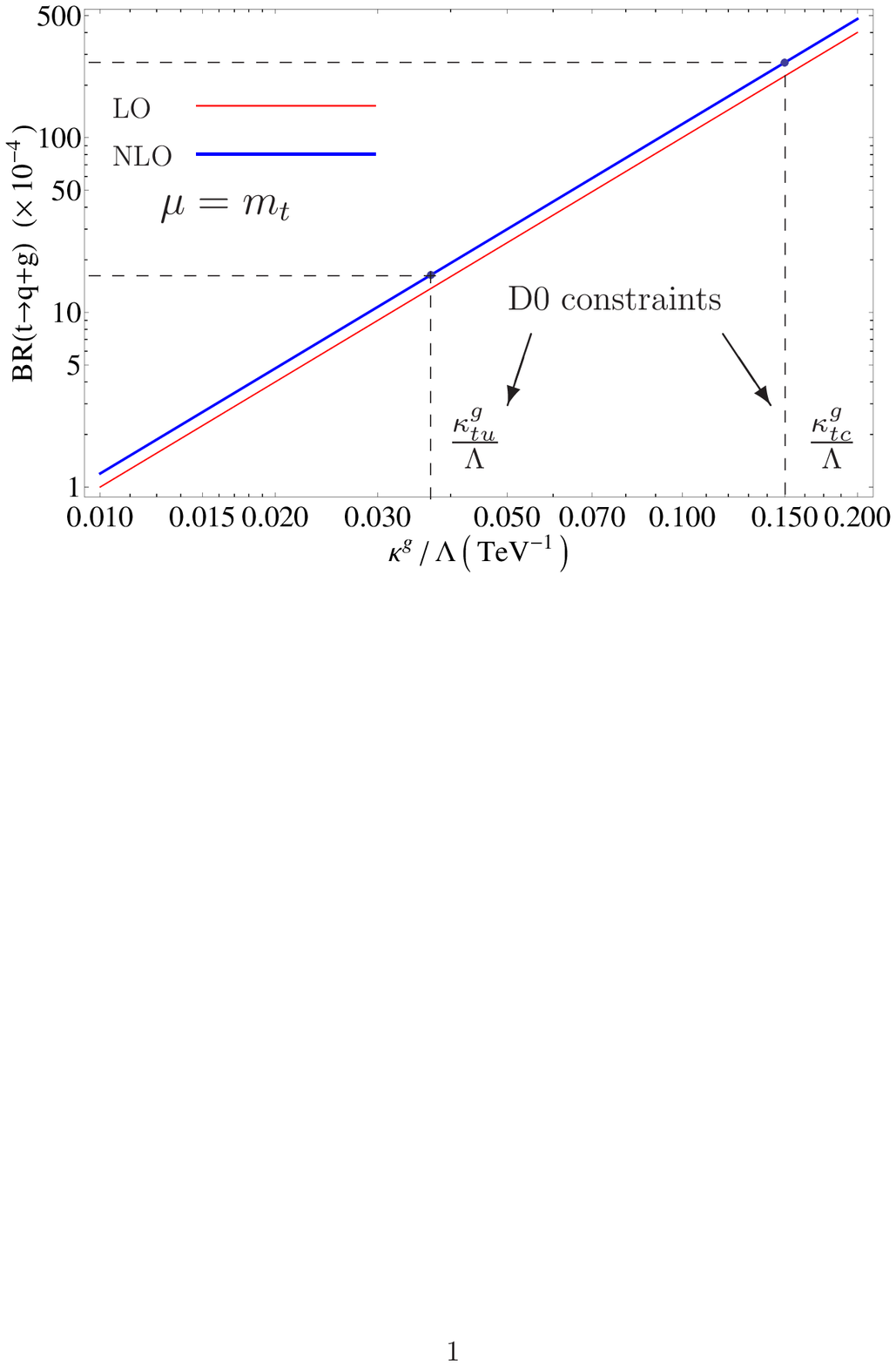}}
\caption{\label{BRvsKappa} Branching ratio of $t\rightarrow q+g$ as
function of $\kappa^g_{tq}/\Lambda$. The D0 upper limits from
Ref.~\cite{Abazov:2007ev} is shown.}
\end{figure}
For convenience, we show the branching ratio of $t\rightarrow q+g$
as function of $\kappa^g_{tq}/\Lambda$ in Fig.~\ref{BRvsKappa},
where we have set $\mu=m_t$ as before. Using the upper limits
measured by the D0 Collaboration at the
Tevatron~\cite{Abazov:2007ev}, we get the following results:
\begin{eqnarray*}
\frac{\kappa^g_{tc}}{\Lambda}<0.15\,\mathrm{TeV}^{-1}&\Rightarrow&
BR(t\rightarrow c+g)<2.69\times10^{-2},\\
\frac{\kappa^g_{tu}}{\Lambda}<0.037\,\mathrm{TeV}^{-1}&\Rightarrow&
BR(t\rightarrow u+g)<1.64\times10^{-3}.
\end{eqnarray*}

\begin{figure}
\scalebox{0.63}{\includegraphics*[120,415][485,650]{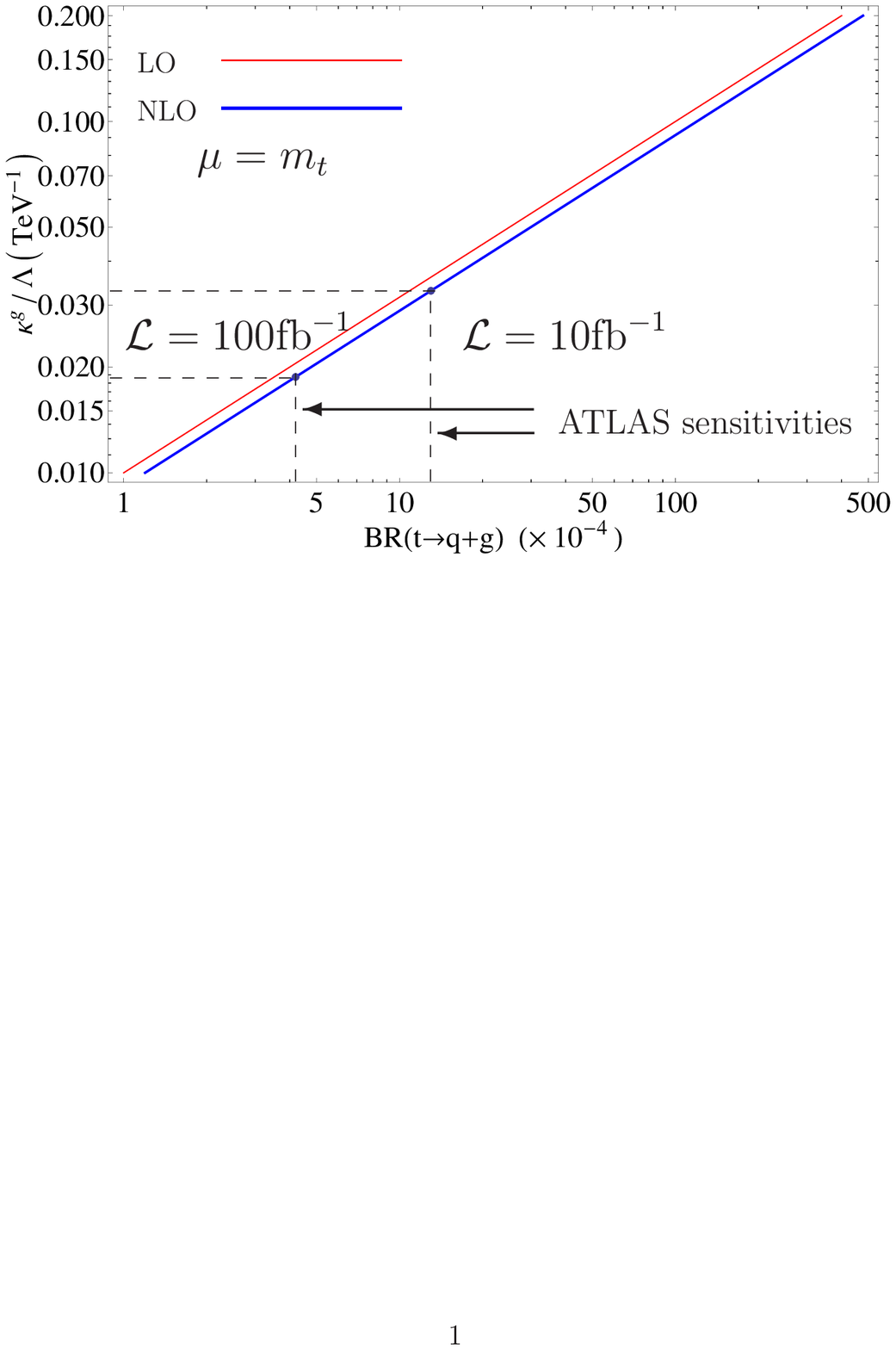}}
\caption{\label{KappavsBR} Coupling $\kappa^g_{tq}/\Lambda$ as
function of branching ratio. The ATLAS sensitivities from
Ref.~\cite{Carvalho:2007yi} is shown.}
\end{figure}
Following the analysis in Ref.~\cite{Carvalho:2007yi}, we plot the
coupling $\kappa^g_{tq}/\Lambda$ as a function of the branching
ratio in Fig.~\ref{KappavsBR} where the ATLAS sensitivities for the
two different expected integrated luminosities are also exhibited.
From Fig.~\ref{KappavsBR}, it is palpable that the NLO prediction
improves the sensitivities of the LHC experiments to measuring the
top quark FCNC couplings. With an integrated luminosity of
$10\,\mathrm{fb}^{-1}$, the ATLAS experiment sensitivities can be
translated to the following relations on FCNC couplings:
\begin{eqnarray*}
BR(t\rightarrow q+g)\geq1.3\times10^{-3}&\Rightarrow&
\frac{\kappa^g_{tq}}{\Lambda}\geq0.033\,\mathrm{TeV}^{-1},\\
BR(t\rightarrow q+\gamma)\geq4.1\times10^{-5}&\Rightarrow&
\frac{\kappa^\gamma_{tq}}{\Lambda}\geq0.028\,\mathrm{TeV}^{-1},\\
BR(t\rightarrow q+Z)\geq3.1\times10^{-4}&\Rightarrow&
\frac{\kappa^Z_{tq}}{\Lambda}\geq0.082\,\mathrm{TeV}^{-1},
\end{eqnarray*}
and with an integrated luminosity of $100\,\mathrm{fb}^{-1}$, they
can be translated to the more stringent relations:
\begin{eqnarray*}
BR(t\rightarrow q+g)\geq4.2\times10^{-4}&\Rightarrow&
\frac{\kappa^g_{tq}}{\Lambda}\geq0.019\,\mathrm{TeV}^{-1},\\
BR(t\rightarrow q+\gamma)\geq1.2\times10^{-5}&\Rightarrow&
\frac{\kappa^\gamma_{tq}}{\Lambda}\geq0.015\,\mathrm{TeV}^{-1},\\
BR(t\rightarrow q+Z)\geq6.1\times10^{-5}&\Rightarrow&
\frac{\kappa^Z_{tq}}{\Lambda}\geq0.036\,\mathrm{TeV}^{-1}.
\end{eqnarray*}

\begin{figure}
\scalebox{0.63}{\includegraphics*[120,415][490,655]{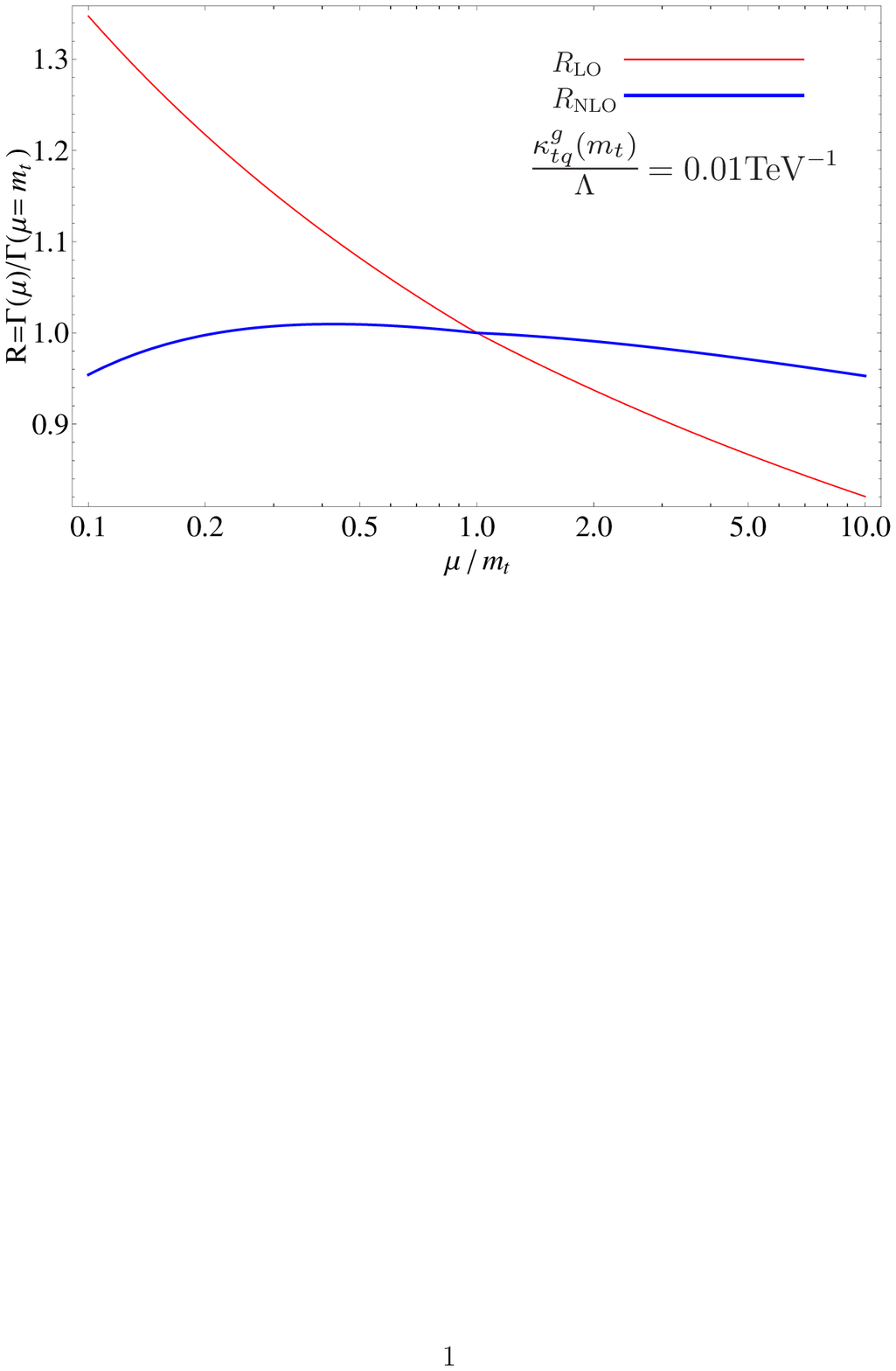}}
\caption{\label{RvsScale}Ratio $R$ as function of the
renormalization scale $\mu$. }
\end{figure}
Lastly, we illustrate the fact that the NLO prediction reduces the
theoretical uncertainty in its prediction on the decay branching
ratios and partial decay widths of the top quark. We define
$R_{\mathrm{LO}}(\mu)=\Gamma_0(\mu)/\Gamma_0(\mu=m_t)$ and
$R_{\mathrm{NLO}}(\mu)=\Gamma(\mu)/\Gamma(\mu=m_t)$, and show the
value of $R(\mu)$ as functions of $\mu$ for $t\rightarrow q+g$ in
Fig.~\ref{RvsScale}. It shows that the theoretical uncertainty from
the renormalization scale dependence can be largely reduced to a
couple of percent once the NLO calculation is taken into account.


In conclusion, in order to achieve consistent studies for both the
top quark production and decay via FCNC couplings, we have
calculated the NLO QCD corrections to the three decay modes of the
top quark induced by model-independent FCNC couplings of dimension 5
operators. For $t\rightarrow q+g$, the NLO results increase the
experimental sensitivity to the anomalous couplings. Our results
show that the NLO QCD corrections enhance the LO branching ratio
about $20\%$. Moreover, the NLO QCD corrections vastly reduce the
dependence on the renormalization scale, which leads to increased
confidence in our theoretical predictions based on these results.
For $t\rightarrow q+\gamma$ and $t\rightarrow q+Z$, the NLO
corrections are minuscule in branching ratios, albeit they can
decrease the LO widths by about $9\%$ and $7\%$, respectively.

C. S. Li would like to thank Wolfgang Wagner for useful
communication that leads to this study. This work was supported in
part by the National Natural Science Foundation of China, under
Grants No.10721063, No.10575001 and No.10635030.
CPY was supported in part by the U.S. National Science Foundation
under award PHY-0555545.

\end{document}